\documentclass[12pt,twoside]{article}
\usepackage{fleqn,espcrc1}
\usepackage{amssymb}
\usepackage{graphicx}
\usepackage[figuresright]{rotating}

\newcommand{\AmS}{{\protect\the\textfont2
  A\kern-.1667em\lower.5ex\hbox{M}\kern-.125emS}}

\title{Two Meson Scattering Amplitudes and their Resonances from Chiral 
Symmetry and the N/D Method\thanks{Talk given at PANIC99, Uppsala (Sweden), 
June 10-16, 1999.}}
\author{J.A. Oller and E. Oset
\address{Departamento de F\'{\i}sica Te\'orica \\
Universidad de Valencia, 46100 Burjasot (Valencia), Spain} 
\thanks{Work partially supported by DGICYT under contract PB96-0753 and by the
EEC-TMR Program-Contrac No. ERBFMRX-CT98-0169. J.A.O. acknowledges financial
support from the Generalitat Valenciana.}}
       
\begin{document}

% typeset front matter
\maketitle

\begin{abstract}
We study the vector and scalar meson-meson amplitudes up to
$\sqrt{s}\lesssim 1.4$ GeV and their associated spectroscopy. The study has been
done considering jointly the N/D method, Chiral Symmetry and implications from
large $N_c$ QCD \cite{nd}. The N/D method provides us with the way to unitarize
the tree level amplitudes constructed in agreement with Chiral Symmetry and its
breaking (explicit and spontaneous). These amplitudes are calculated making use
of the lowest order Chiral Perturbation Theory ($\chi PT$) Lagrangians 
\cite{xpt} and the exchanges of resonances compatible with Chiral Symmetry as 
given in \cite{grupo}. On the other hand the large $N_c$ considerations allow 
us to distinguish between elementary (as elementary as the pions, for instance) and
compound (meson-meson) states. Making use of this formalism one observes that
the $\sigma$, $\kappa$ and $a_0(980)$ resonances are meson-meson states
originating from the unitarization of the lowest order $\chi PT$ amplitudes. On
the other hand, the $f_0(980)$ is a combination of a strong S-wave meson-meson
unitarity effect and of a preexisting singlet resonance with a mass around 1
GeV.

For the vector resonances we reproduce the well known features of Vector Meson
Dominance (VMD) and the KSFR relation \cite{ksfr}. The much more important role
that unitarity plays in the scalar sector as compared with the vector one is
also stressed. In particular, it is clear from our study that the $\sigma$ and
the $\rho$ resonances are states completely different in nature. The $\sigma$ is
a two pion resonance originated from the interaction of the pions whereas the
$\rho$ is a state as elementary as the pions themselves whose origin has nothing
to do with the interaction between pions.
\end{abstract}

\section{Introduction} \label{sec:intro}

	$\chi PT$ \cite{xpt} can be supplied with the exchange of explicit 
resonance fields \cite{grupo}. In doing this, a resummation up to an infinite
order in the chiral expansion can be achieved from the
expansion of the bare propagator of a resonance. In fact, at 
$\mathcal{O}(p^4)$, it is seen \cite{grupo} that the $L_i$ counterterms of 
$\chi PT$ are saturated
by the exchange of the resonances. However, the amplitudes that can be
built directly from $\chi PT$ at $\mathcal{O}(p^4)$ plus resonance exchanges as
in \cite{UMR}, need a unitarization procedure in order to compare directly
with experimental data (phase shifts, inelasticities...) for the different
energy regions, in particular, around the resonance masses. This is one of the
aims of the present study.

	On the other hand, it is well known that the scalar sector is much more
controversial than the vector or tensor ones. In the latter case, the
associated spectroscopy can be understood in terms of first principles coming
directly from QCD as Chiral
Symmetry and Large $N_c$ plus unitarity, once we admit VMD as dictated by
phenomenlogy. We try to make use in this work of the same principles than before,
that is, Chiral Symmetry, Large $N_c$ and unitarity in coupled channels, in
order to study the scalar resonant channels.

\section{Formalism} \label{sec:form}
We consider the influence of the unphysical cuts perturbatively.
In this we take the zero order approach, that is, we neglect it. In 
\cite{nd} we make estimations
of the unphysical cuts contribution and find them to be only a few per
cent of our final amplitudes. In \cite{jamin} it is discussed how 
to include them up to one loop calculated at $\mathcal{O}(p^4)$. 

When taking into account the N/D method \cite{ndmethod} the most general
structure that an elastic partial wave amplitude, $\hbox{T}_L$, has when the 
unphysical cuts are neglected, is \cite{nd}:

\begin{eqnarray}
\label{fin/d}
\hbox{T}'_L(s)&=&\frac{1}{\hbox{D}_L(s)}\nonumber\\
\hbox{D}_L(s)&=&-\frac{(s-s_0)^{L+1}}{\pi}\int_{s_{th}}^\infty ds' \frac
{\nu(s')^L \rho (s')}{(s'-s)(s'-s_0)^{L+1}}+\sum_{m=0}^L a_m s^m+
\sum_i^{M_L} \frac{R_i}{s-s_i}
\end{eqnarray}
where $\hbox{T}'_L(s)=\hbox{T}_L(s)/\nu^L$, $\nu^L=p^{2L}$ with $p$ the
center mass three momentum of the particles and $s_0$ is the subtraction point.

In Large $N_c$ counting rules $\hbox{T}_L\rightarrow 1/N_c$ and hence
$\hbox{D}_L\rightarrow N_c$. We then split the $\hbox{D}_L$ in two parts:

\begin{eqnarray}
\label{split}
\hbox{D}_L^{\infty}&=&\sum_{m=0}^L a_m^L s^m+
\sum_i^{M_L} \frac{R_i}{s-s_i} \rightarrow \mathcal{O}(N_c) \nonumber \\
 g(s)&=&-\frac{(s-s_0)^{L+1}}{\pi}\int_{s_{th}}^\infty ds' \frac
{\nu(s')^L \rho (s')}{(s'-s)(s'-s_0)^{L+1}}+\sum_{m=0}^L a_m^{SL} s^m
\rightarrow \mathcal{O}(1)
\end{eqnarray}
where $a_m^{L(SL)}$ is the $\mathcal{O}(N_c)$($\mathcal{O}(1)$) of the
coefficient $a_m$.

In the large $N_c$ limit a partial wave amplitude is given \cite{witen} by local
plus pole terms. The local terms are taken from the lowest order $\chi PT$
amplitudes and the resonant ones from the exchange of resonances as given in
\cite{grupo}. In doing this, we are assuming the concept of Meson Resonance
Dominance, in the sense that all the local terms of order higher than two in
the $\chi PT$ expansion are saturated by the resonance exchanges. This was shown
to be case at $\mathcal{O}(p^4)$ in \cite{grupo}. In this way

\begin{equation}
\label{large}
\hbox{T}_L^{\infty}=\frac{1}{\hbox{D}_L^\infty}=T^{(2)}+T^R
\end{equation}
where $T^{(2)}$ refers to the lowest order $\chi PT$ amplitude and $T^R$ to the
resonance exchange amplitudes. In \cite{nd} it is proved that
$\hbox{D}_L^\infty$ has enough room to accommodate $T^{(2)}+T^R$ as given 
above. Thus, our final formula will be:

\begin{equation}
\label{final}
\hbox{T}_L=p^L \left(\left[T^{(2)}+T^R\right]^{-1}+g(s)\right)^{-1} p^L
\end{equation}

The extension of the former formula to the coupled channel case is
straightforward in a matrix formalism. Then, we have: $T^{(2)}$, 
$T^R$ $\rightarrow$ $T^{(2)}_{ij}$, $T^R_{ij}$ and $g(s)$ and $p$ are 
now diagonal matrices \cite{nd}.

\section{Vectors}

Making use of the former formalism we study simultaneously the $\rho$ and 
$K^*$ resonances \cite{nd}. We obtain the expected leading behaviours of their 
bare poles and the KSFR \cite{ksfr} value of the coupling with a relative 
deviation of only a $6 \%$. The former conclusions are obtained after fitting
the elastic P-wave $\pi\pi$ and $K\pi$ phase shifts in terms of two parameters:
the subtraction constant $a^{SL}$ present in $g(s)$ and the coupling of the
vector octet of resonances.

\section{Scalars}

We study the $SU(3)$ connected meson-meson S-waves with $I=0$, 1 and 2,
where $I$ refers to the isospin. Contrary with the former vector channels, which
are essentially elastic, for the proposed scalar channels one has to take into
account the effect of coupling channels. We include the following channels:
\begin{equation}
\label{channels}
\begin{array}{ll}
I=0 & \pi\pi(1), \; K\bar{K}(2), \; \eta\eta(3)\\
I=1 & \pi\eta(1), \;  K\bar{K}(2)\\
I=1/2& K\pi(1), \; K\eta(2) 
\end{array}
\end{equation}

These are the most relevant channels up to $\sqrt{s} \approx 1.2$ GeV. For
energies higher than these other channels become increasingly more important as
four pions for $I=0$ or $K\eta'$ $I=1/2$. Thus, a proper study of the set of 
resonances 
that we find around 1.4 GeV would require the inclusion of those relevant 
channels \cite{jamin}.

We fit, up to $\sqrt{s}\lesssim 1.4$ GeV, the following data: $I=0$ 
elastic $\pi\pi$ phase shifts, $I=0$ $K\bar{K}\rightarrow \pi\pi$ phase shifts,
$\frac{1-\eta^2_{00}}{4}$ where $\eta_{00}$ is the inelasticity with $I=0$,
$I=1/2$ elastic $K\pi$ phase shifts and a mass distribution for $I=1$ around the
$a_0(980)$ resonance.

In a first glance at the PDG \cite{pdg} one can think that there should be at
least two scalar nonets, one with a mass below 1 GeV and another around 1.4 GeV. 
We first include two scalar nonets but then the fit gives a remarkable feature. 
The octet around 1 GeV has vanishing coupling constants and the same occurs with
 the singlet around 1.4 GeV. As a consequence, one can reproduce the scalar 
 data with
only a singlet around 1 GeV and an octet around 1.4 GeV. The $\chi^2$ per degree
of freedom obtained is almost 1 with 188 experimental points.

From the point of view of the resonance content of our amplitudes, the
explicitly included octet with a mass about 1.4 GeV evolves to give 
rise to poles with masses very close to those of the
$K^*_0(1430)$, $a_0(1450)$ and $f_0(1500)$ resonances. In turn, the singlet 
around 1 GeV evolves to the physical pole of the $f_0(980)$. However, together
with the former poles we also find other ones which do not originate from any
preexisting resonance ($T^R=0$). They 
are meson-meson resonances originating from the
unitarization of the lowest order $\chi PT$ amplitudes. On the other hand, since
loops are suppressed in large $N_c$ these resonances disappear for
$N_c\rightarrow \infty$. They correspond to the $\sigma(500)$, $a_0(980)$,
$\kappa$ and a strong contribution to the physical $f_0(980)$ resonance. Thus,
the $f_0(980)$ resonance results from two effects: a preexisting resonance
around 1 GeV and a strong $K\bar{K}$ threshold effect.

\section{Conclusions}

We have presented a systematic procedure to unitarize the tree level amplitudes
coming from lowest order $\chi PT$ and the explicit exchange of resonance
fields. We have used this method to study the vector and scalar resonances. For
the vectors, we reproduce the well known features of VMD and the KSFR value for the
coupling of the vector resonances $\rho$ and $K^*$. For the controversial scalar
channel, the situation is more complicated. After reproducing a large amount of
experimental data, we have observed two sets of
resonances. Those resonances preexisting to the unitarization: one octet around 1.4 GeV 
and a singlet around 1 GeV that evolves to the physical
$f_0(980)$ resonance. The other set corresponds to meson-meson resonances with a
mass $\lesssim$1 GeV: $\sigma$, $a_0(980)$, $\kappa$ and a strong contribution
to the $f_0(980)$ from the $K\bar{K}$ threshold. This set of resonances forms a
nonet and in fact when we go in our formalism to the $SU(3)$ limit they form an
octet of degenerate resonances plus a singlet.

\end{document}